\documentclass[review,12pt,3p]{elsarticle}
\usepackage{graphicx}
\usepackage{caption}
\usepackage{subcaption}
\usepackage{amssymb}
\usepackage{amsmath}
\usepackage{amsthm}  %% for the mathematical proof of the hypotheses below
\usepackage{hyperref}

\newtheorem{hypothesis}{Hypothesis}

\journal{Measurement} %% or \journal{Physics Letters A} 

\begin{document}

\begin{frontmatter}

\title{Thermal tuning of light-emitting diode wavelength as an implication of the Varshni equation}

\author{Richard O. Ocaya}
\ead{ocayaro@ufs.ac.za}
\address{Department of Physics, University of the Free State, South Africa}

\begin{abstract}
In this article, we show that the variation of the wavelength of a non-pumped light-emitting diode (LED) is practically linear with temperature, a novel consequence of Varshni's widely accepted empirical expression. This formula models the bandgap variation for most semiconductors from 0K to their high-temperature limits, subject to fitting parameters. Therefore, we suggest an external thermal mechanism that can be used to tune the wavelength of a constant-current biased LED and also to stabilize its wavelength. Furthermore, we demonstrate the approach on published AlN data and show that the fitting parameters follow trivially. In addition, we suggest a novel method to characterize semiconductors. Finally, we present the results of experimental measurements on several commercial LEDs.
\end{abstract}


\end{frontmatter}
% \linenumbers

\section{Introduction}

Although LEDs have been in existence in some form or other for several decades \cite{zheludev2007life,bourget2008introduction}, they have become more prominent and ubiquitous over the last few years, driven by the demand for efficient (large, low-power, high-brightness) displays on mobile devices. The research interest in this field is likely to grow as new combinations of LED-suited materials continue to be found. LEDs, both new and traditional, rely on semiconductor material in which a bound electron can be electrically coerced to transit a characteristic energy gap and then to undergo a radiative recombination into a bound state. The radiation is at the characteristic wavelength of the material. For radiation to occur efficiently the transition must not involve an exchange of momentum with the host lattice. Otherwise, the transitions would involve phonons (heat), causing low photon absorption and radiation efficiency.  It is for this reason that silicon and germanium, which are indirect bandgap materials, are not suited for LEDs. In traditional semiconductors this energy gap is the so-called bandgap. In the newer organic semiconductors, an energetic process akin to the bandgap arises due to the presence of the highest-occupied and lowest-unoccupied molecular orbitals (HOMO/LUMO) \cite{aihara1999weighted,kim2013energy}.  Notwithstanding the interest in the newer organic semiconductors, traditional compound semiconductors are still important because they possess higher mobilities and other advantages over elemental semiconductors, silicon especially. They employ mostly Group II-V in controlled alloys of elements such as gallium, nitrogen, arsenic, indium, aluminium, phosphorus and others. The multitude of resultant direct bandgap materials, which can range from binary to quaternary alloys, have diverse properties and applications, such as LEDs. The bandgaps can be controlled by adjusting the elemental stoichiometric ratios. As with bandgaps, HOMO/LUMO energy gaps can also be engineered \cite{kaur2008substituent}.

The precise value and behaviour of the effective bandgap of a material with respect to external influences, particularly temperature, are vitally important. However, it has been shown that certain, carefully grown compound semiconductors have bandgaps that are insensitive to temperature \cite{oe1996proposal,asahi2017temperature}. Such materials are suitable for wavelength-stable laser diodes. Sze and Ng \cite{sze2006physics} mention that the emission wavelength of the Fabry-Perot laser follows the bandgap temperature variation. They present the specific case of the PbTe/Pb$_{1-x}$Sn$_x$Te laser with a temperature-tunable wavelength. The quantification of the behavior for the majority of materials has mostly relied on photoluminescence spectral analysis over temperature \cite{Guo_1994,walukiewicz2004optical}. In the literature, Varshini's relation (VR) has been fitted to many semiconductors, both elemental e.g. Si and Ge, and compound \cite{bhowmick2019mathematical,krustok2019observation,xu2019fast,isik2020temperature}. Recently, it was demonstrated for nanowires \cite{xu2019fast}, in the context of wavelength tuning, quantum dots \cite{magaryan2019analysis}, and high-voltage and high-temperature (wide bandgap) devices \cite{raynaud2010comparison}. Admittedly, the theoretical basis of the Varshini equation is weak because of its complete reliance on empirical data \cite{vainshtein1999applicability,o1991temperature,fang1990photoluminescence,manoogian1979determination}. Even the heavily referenced work of Sze and Ng {\em assumes} a linear reduction of the 0K bandgap with temperature for a semiconductor diode \cite{sze2006physics}. There are important cases in the literature where a crucial fitting constant in the Varshini equation is negative, thereby going against the norm. However, several proposals have been made over the years for the bandgap-temperature variation such as the Manoogian-Leclerc relation \cite{manoogian1979determination}, and others \cite{o1991temperature}. These newer models approach the decrease in bandgap by considering lattice expansion, and the acoustic and optical oscillations due to electron-phonon coupling within the context of Bose-Einstein statistics as complex functions of temperature. 

In this paper, we assume the validity of VR as the starting point. Then, we present and prove two hypotheses under simple assumptions. We devise a basic experimental setup to vary the temperature of the LED under test while recording its wavelength. The LED is biased at a constant current that produces negligible self-heating. For these, several discrete LEDs were used. Finally, we present the results of our experiments to validate the hypotheses. To the best of our knowledge, this approach has not hitherto been attempted for LEDs. It holds the promise to extend the wavelengths of LED sources through externally-exerted temperature control in a novel way. In addition, it adds a new tool into the arsenal to characterize LEDs, particularly new ones. The variation of the wavelength, even for the restricted temperature range of our own experiments, is sufficiently large, particularly for the green LED, to be visible to the naked eye as a reversible color changes. Most elementary experiments show that when LEDs are driven at higher voltages or currents there is an apparent visible color change towards longer wavelengths (i.e. a red shift). If this over-driving of the LED is continued, it typically culminates in the destruction of the LED. This work also explains, in passing, this observed color change as being due to LED self-heating rather than the consequence of the creation of new energy states by the applied electric field.

\section{Theory}

The Varshni equation relates the bandgap $E_g$ of a material at temperature $T$ to its 0K bandgap $E_{g,0}$ by
\begin{equation}\label{eqn:varshni}
E_g(T)=E_{g,0}-\frac{\alpha T^2}{\beta+T},    
\end{equation}
where $\alpha$ and $\beta$ are semiconductor-dependent fitting constants. We note that, typically, $\alpha$$\equiv$ eV/K and $\beta$$\equiv$ K. For LEDs, it is convenient to work in units of nanometer (nm) and Kelvin (K). Equation (\ref{eqn:varshni}) can then be converted to these units through a Planck's equation factor, $\varepsilon$=$10^9hc/\lambda$$\approx$1.988$\times$$10^{-16}$ (eV.nm) i.e.
\begin{equation}\label{eqn:varshni2}
    \frac{\varepsilon}{\lambda(T)}=\frac{\varepsilon}{\lambda_0}-\frac{e\alpha T^2}{\beta+T}
\end{equation}
where $h$=6.626$\times$ $10^{-34}$ J.s is Planck's constant, $c$=3.0$\times$ $10^8$ m/s is the speed of light. In addition, $e$ is electronic charge, $10^9$ is the nanometer conversion factor and $\lambda_0$ is taken as the emitted LED wavelength at 0K. Equation (\ref{eqn:varshni2}) then becomes 
\begin{eqnarray}
    \nonumber\lambda(T)&=&\frac{\varepsilon(\beta+T)\lambda_0}{\varepsilon(\beta+T)-e\alpha\lambda_0 T^2}\\
    &=&\frac{\lambda_0}{1-\frac{\delta T^2}{\beta+T}},\label{eqn:varshni3}
\end{eqnarray}
where $\delta$=$e\alpha\lambda_0/\varepsilon$ in units of /K.
We now take a detour at this point to illustrate the typical magnitude of $\delta$ using literature data, for instance \cite{wu2003temperature,vainshtein1999applicability}. Table \ref{tab:materials} reproduces published literature and calculates $\delta$ for each material. The values of $\lambda_0$ are calculated from $E_{g,0}$ and are included for the sake of completeness.
\begin{table}[!htb]
\centering
\caption{\label{tab:materials}Varshni's parameters of different materials from the literature based on experiments.}
\begin{tabular}{@{}lllllll}
\hline
 & $E_{g,0}$ & $\lambda_0$& $\alpha$ & $\beta$ & $\delta$ &  \\
Material & (eV) & (nm) & $\times$10$^{-4}$(eV/K) & (K) & $\times$10$^{-4}$(/K) & {\small Ref.}\\
\hline
AlN & 6.23 & 199.1 & 5.93 & 600 & 0.96 & {\small \cite{Guo_1994,wu2003temperature}} \\
%% GaN & 3.51 & 353.5 & 9.09 & 830 & 2.59 & \\
Si & 1.17 & 1060.4 & 7.02 & 1108 & 6.00 & {\small \cite{vainshtein1999applicability,bart}}\\
Ge & 0.74 & 1676.6 & 4.56 & 210 & 6.16 & {\small \cite{vainshtein1999applicability,bart}}\\
ZnSe & 2.82 & 440.0 & 5.78 & 175 & 2.05 & {\small \cite{vainshtein1999applicability}}  \\
GaAs & 1.52 & 816.2 & 8.87 & 572 & 5.84 & {\small \cite{vainshtein1999applicability}} \\
GaP & 2.32 & 534.8 & 6.86 & 576 & 2.96 & {\small \cite{vainshtein1999applicability}} \\
InAs & 1.42 & 837.7 & 3.16 & 93 & 2.23 & {\small \cite{vainshtein1999applicability}} \\
\hline
\end{tabular}
\end{table}
The table shows that $\delta$ is typically very small for most semiconductors. This is true also for materials used in LEDs. We then return to the matter at hand with the knowledge that typically $\delta<<1$. We then present and prove that for a LED wavelength tuning by adjusting its temperature is possible.

\subsection{Hypotheses}
We present and prove the following hypotheses that establish the theoretical aspect of this article. The theory is then tested in a subsequent section to determine the parameters of the Varshni relation.
\begin{hypothesis}\label{hypo:one}
Let Equation (\ref{eqn:varshni}) be valid for a practical LED material on a given temperature range. Then, provided that $\delta$ $<<1$, the temperature variation of the emitted wavelength can be written
\begin{equation}\label{eqn:wavelength}
    \lambda(T)=\lambda_0\left(1+\frac{\delta T^2}{\beta+T}\right).
\end{equation}
\end{hypothesis}
\begin{proof}
Equation (\ref{eqn:varshni3}) can be written as
\begin{equation}\label{eqn:wavelength2}
    \lambda(x)=\frac{\lambda_0}{1-x},\quad\textrm{where}~x=\frac{\delta T^2}{\beta+T}.
\end{equation}
This equation can be expanded using the binomial theorem into a convergent infinite series i.e.
\begin{equation}\label{eqn:series}
\frac{1}{1-x}=\sum_{n=0}^\infty~x^n,\qquad\textrm{provided that}~|x|<1.
\end{equation}
Thus, for Equation (\ref{eqn:varshni3}) to converge in the variable $T$, the necessary condition is that
\begin{eqnarray}
   \nonumber -1<\frac{\delta T^2}{\beta+T}<1&&\\
    \implies\textrm{either}~(1): -\left(\frac{\beta+T}{T^2}\right)<\delta\qquad \textrm{or}~ && (2): \delta<\frac{\beta+T}{T^2}.\label{eqn:limiting}
\end{eqnarray}
We discard option $(1)$ in Equation (\ref{eqn:limiting}) since $\{\beta,T\}>0$K would give $\delta<0$, which is clearly not possible, as illustrated in Table \ref{tab:materials}. Then, quadratically solving option $(2)$ for $T\ge 0$K gives the upper limiting temperature:
\begin{equation}
    T_{H}=\frac{1+\sqrt{1+4\delta\beta}}{2\delta}.
\end{equation}
This result indicates that the validity of Equation (\ref{eqn:series}) holds for $\forall T\in[0,T_{H})$. Finally, we note that the right-hand side of Equation (\ref{eqn:series}) can be written as
\begin{equation}\label{eqn:series2}
\sum_{n=0}^\infty~x^n=x^0+x^1+x^2+\cdots=1+x+\mathcal{O}(2)\approx 1+x.
\end{equation}
One can check numerically that, indeed, the series converges with a negligible $\mathcal{O}(2)$ error. Hence, Equation (\ref{eqn:wavelength2}) can be written as
\begin{displaymath}
    \lambda(T)=\lambda_0\left(1+\frac{\delta T^2}{\beta+T}\right),
\end{displaymath}
which completes the proof.
\end{proof}

\begin{hypothesis}\label{hypo:two}
Sufficing that Hypothesis \ref{hypo:one} is valid, then provided that $\beta>>T$, one can write
\begin{equation}
   \lambda(T)=\lambda_0+m T, 
\end{equation}
for a some determinable constant $m$, indicating that the emitted LED wavelength is linear for all purposes over the given temperature range, $T\in[0,T_m)$ where $T_m<<\beta$.   
\end{hypothesis}
\begin{proof}
Applying long division, 
\begin{equation*}
    \frac{\delta T^2}{\beta+T}=\delta T+(-\delta\beta)\left(1-\frac{\beta}{\beta+T}\right)
\end{equation*}
Therefore, Equation (\ref{eqn:wavelength}) can be written
\begin{eqnarray}
    \nonumber\lambda(T)&=&\lambda_0\left[1+{\delta T+(-\delta\beta)\left(1-\frac{\beta}{\beta+T}\right)}\right]\\
     &=&\lambda_0\left[1+{\delta T+(-\delta\beta)+\left(\frac{\delta\beta^2}{\beta+T}\right)}\right].\label{eqn:reduce}
\end{eqnarray}
Then, if $\beta>>T$, with a maximum acceptable value of $T$ being $T_m$, then 
\begin{displaymath}
\frac{\delta\beta^2}{\beta+T_m}\longrightarrow \frac{\delta\beta^2}{\beta}\approx \delta\beta.\label{eqn:reduce2}
\end{displaymath}
Hence Equation (\ref{eqn:reduce}) becomes 
\begin{eqnarray}
    \lambda(T)&=&\lambda_0\left[1+{\delta T}\right],\label{eqn:reduce3}\\
    &=&\lambda_0+mT,\label{eqn:reduce4}
\end{eqnarray}
where $m=\delta\lambda_0$. This completes the proof.
\end{proof}
To summarize, the above hypotheses imply that the variation of wavelength of a LED is, for all practical purposes, linear for sufficiently large values of $\beta$, and over practically realizable temperatures. Such a linear function $\lambda(T)$ has gradient $m$, from which one could determine $\delta$, and thus directly the Varshni fitting coefficient $\alpha$. The second parameter, $\beta$, can then subsequently be calculated. Finally, one can also deduce $\lambda_0$ and then $E_{g,0}$, which are the 0K emission wavelength and bandgap energy, respectively.

\subsection{The Varshni constants from experimental data}
The constant $\delta$ involves the as-yet unknown $\lambda_0$ but can be calculated readily from Equation (\ref{eqn:reduce3}) using the data sets i.e.
\begin{eqnarray}
    \nonumber\lambda_1=\lambda_0\left[1+{\delta T_1}\right],&&\lambda_2=\lambda_0\left[1+{\delta T_2}\right],\\
    \Rightarrow\delta&=&\frac{\lambda_1-\lambda_2}{\lambda_2T_1-\lambda_1T_2}.\label{eqn:deltaval}
\end{eqnarray}
The 0K emission wavelength is then found using either data set, i.e.
\begin{equation}\label{eqn:0Kwavelength}
    \lambda_0=\frac{\lambda_1}{1+\delta T_1}.
\end{equation}
Hence the 0K bandgap can also readily be calculated using the Planck constants. We defined $\delta$=$e\alpha\lambda_0/\varepsilon$. Therefore,
\begin{equation}\label{eqn:delta}
    \alpha=\frac{\delta\varepsilon}{e\lambda_0}.
\end{equation}
Equation (\ref{eqn:varshni3}), which is not an approximation, can be written for two arbitrary {\em wavelength-temperature} data sets ($\lambda_1$,$T_1$) and ($\lambda_2$,$T_2$), such as from experimental measurements:
\begin{equation*}
    \lambda_0=\lambda_1\left(1-\frac{\delta T_1^2}{\beta+T_1}\right)=\lambda_2\left(1-\frac{\delta T_2^2}{\beta+T_2}\right).
\end{equation*}
Hence, 
\begin{equation}\label{eqn:beta}
    \beta^2\left(\frac{\lambda_2}{\lambda_1}-1\right)-\beta\left[\delta(T_2^2-T_1^2)\right]-\left[\delta(T_1T_2)(T_2-T_1)\right]=0.
\end{equation}
This last equation can be solved quadratically for $\beta>0$ since $\delta$ is known. 

To summarize, if the variation of the wavelength $\lambda$ with temperature $T$ could be established as linear with gradient $m$, then it is possible to estimate the Varshni relation for the LED i.e. to determine $\alpha$, $\beta$, and $E_{g,0}$. 

\subsection{Illustration with AlN literature data}
\begin{figure}[h]
 \begin{subfigure}{0.5\textwidth}
\includegraphics[width=\linewidth, height=5cm]{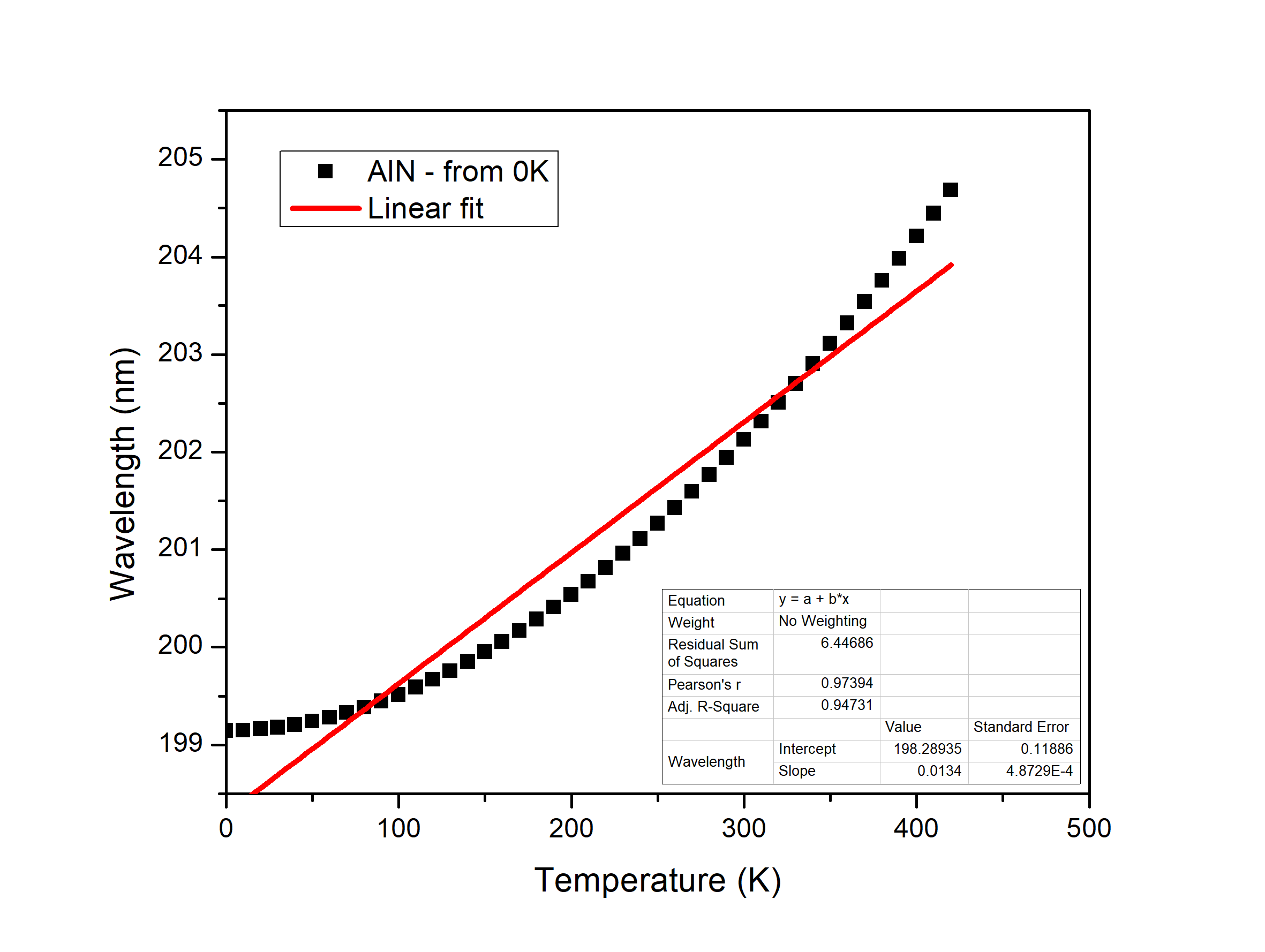} 
\caption{0K - 420K.}
\label{fig:aln0K}
\end{subfigure}
\begin{subfigure}{0.5\textwidth}
\includegraphics[width=\linewidth, height=5cm]{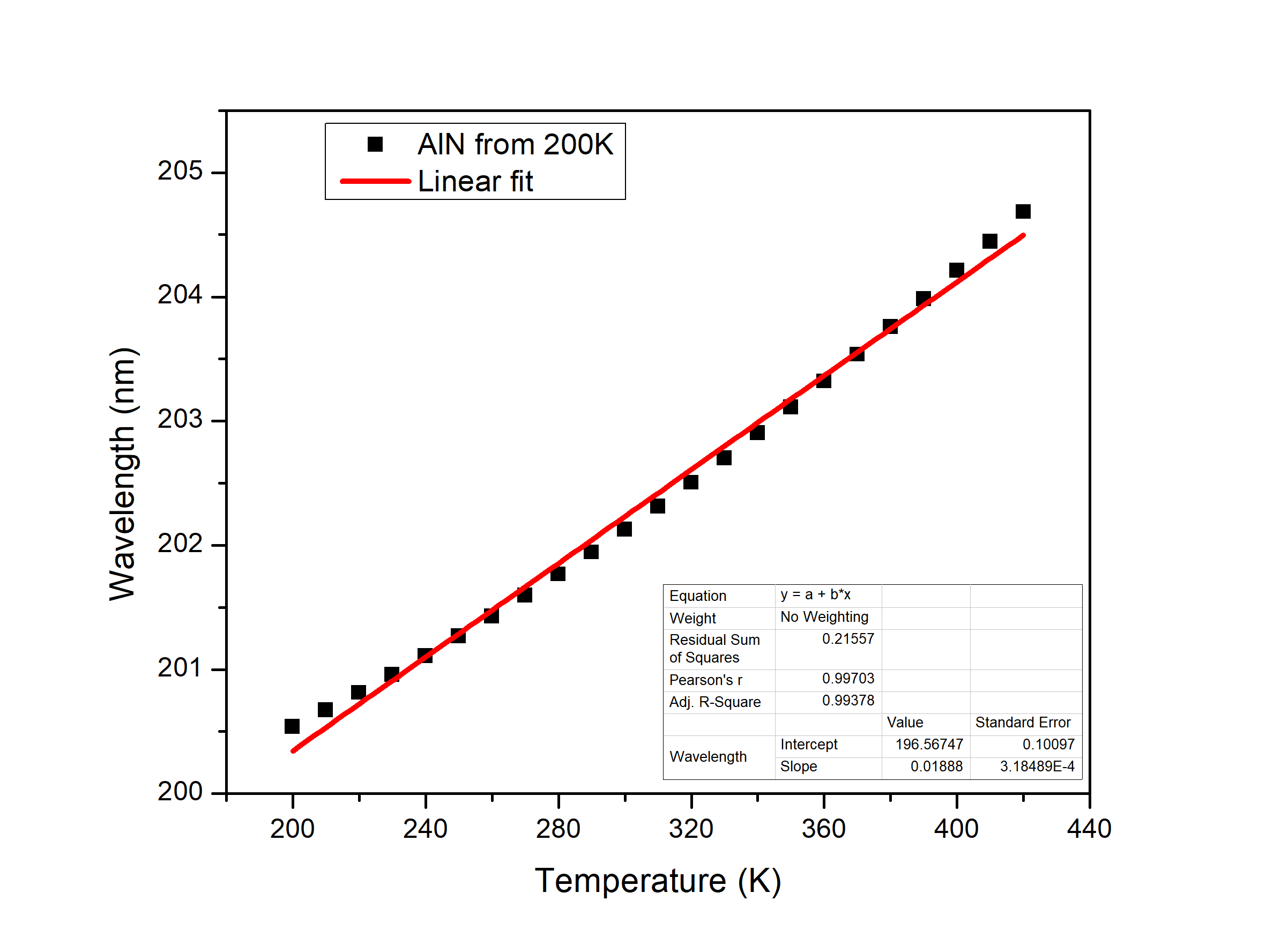}
\caption{200K - 420K.}
\label{fig:aln200K}
\end{subfigure}
 \caption{Calculated wavelength variations of AlN based on published literature data on two temperature ranges.}
\label{fig:alumnitride}
\end{figure}

\begin{figure}[h]
 \begin{subfigure}{0.5\textwidth}
\includegraphics[width=\linewidth, height=5cm]{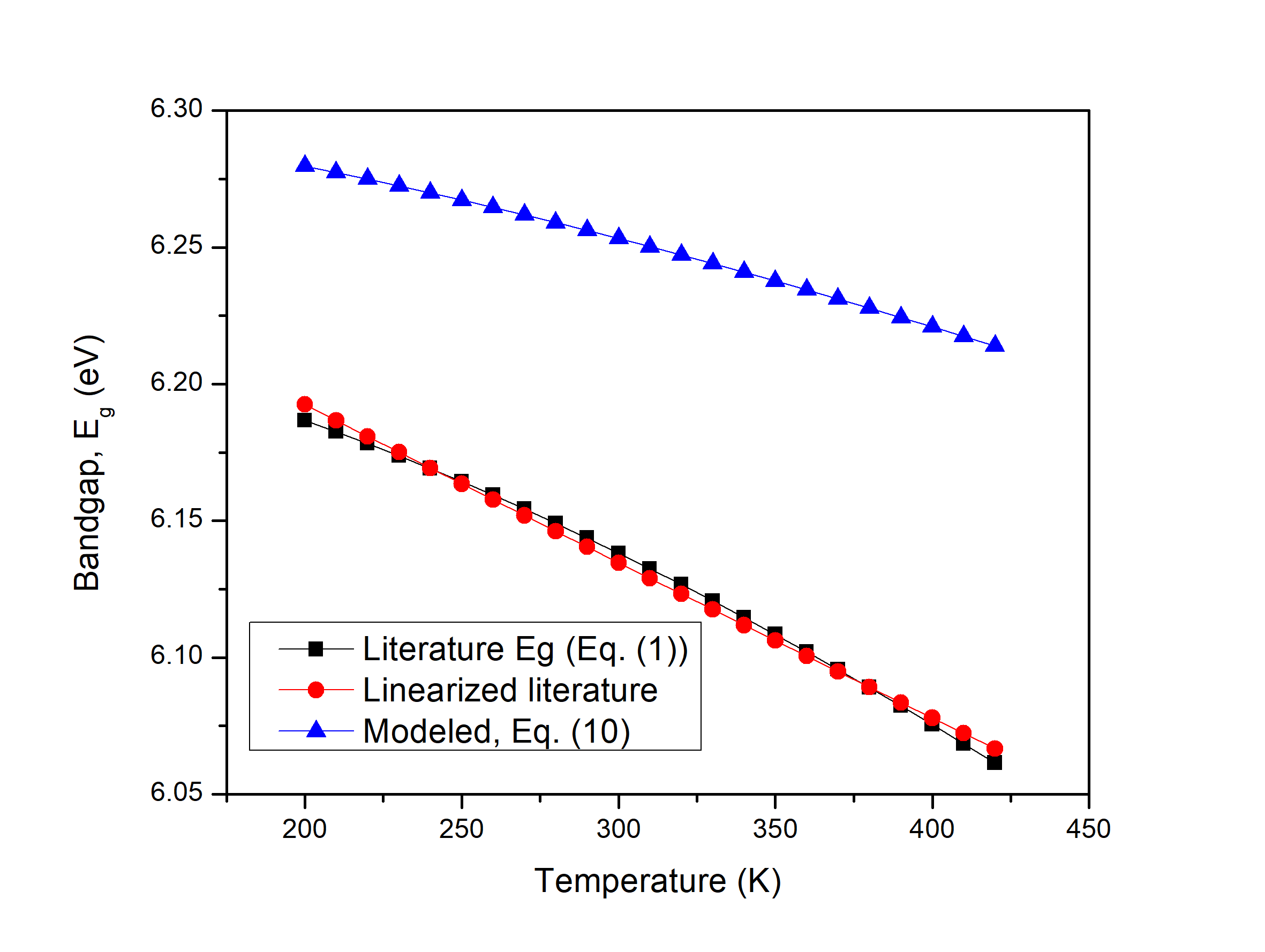} 
\caption{$E_g(T)$}
\label{fig:bandgaps}
\end{subfigure}
\begin{subfigure}{0.5\textwidth}
\includegraphics[width=\linewidth, height=5cm]{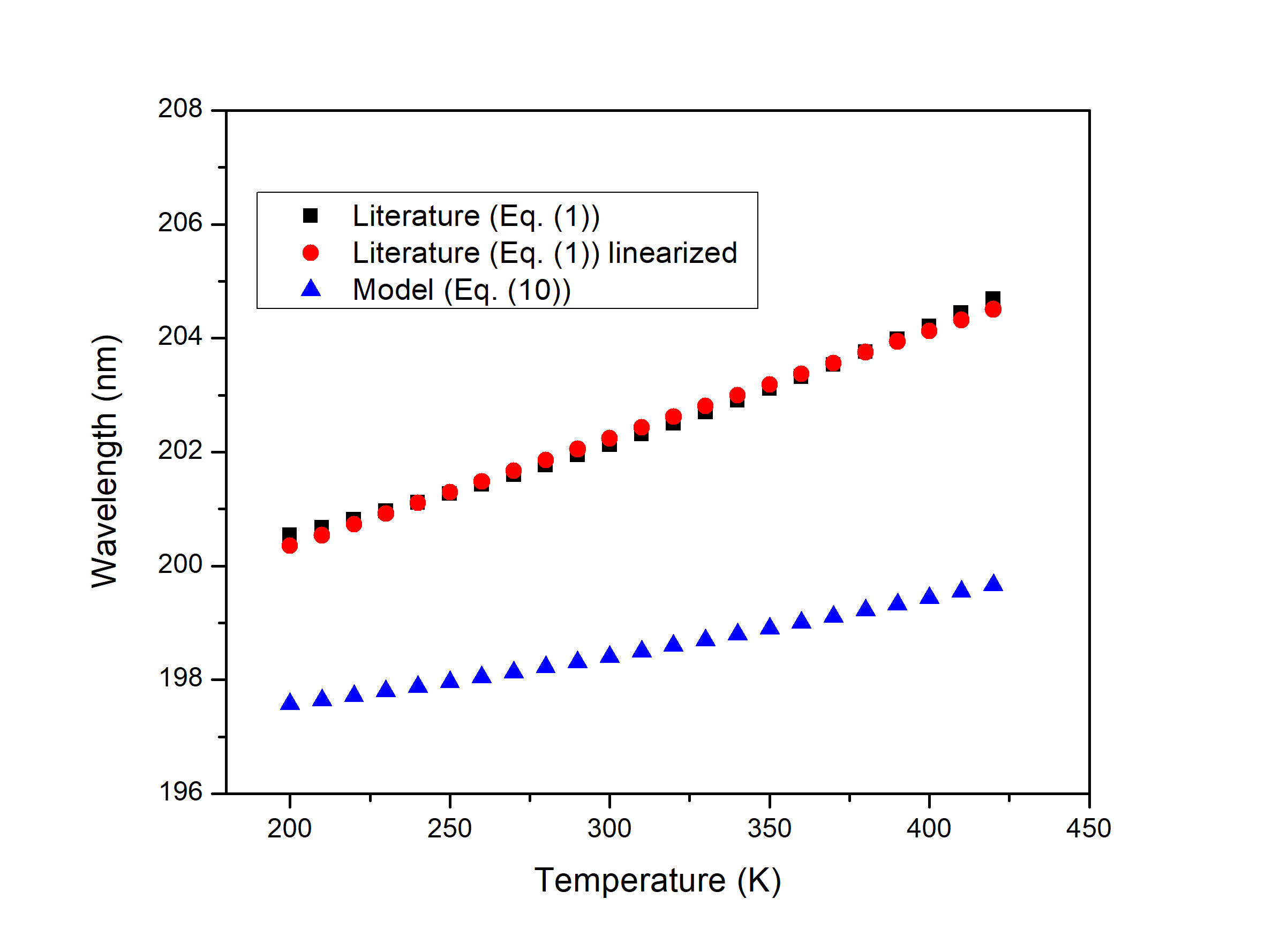}
\caption{$\lambda(T)$}
\label{fig:wavelengths}
\end{subfigure}
 \caption{A comparison of calculated and modeled bandgap and wavelength versus temperature variations for published AlN data. The instantaneous values differ by less than 2\%.}
\label{fig:comparisons}
\end{figure}

In this section, we test the approach using aluminium nitride (AlN) with extrapolated data from the literature \cite{wu2003temperature,vurgaftman2001band,guo1994temperature}. The published data, reproduced in Table \ref{tab:alumnitride}, mentions two sets of values of Varshni's coefficients, $\alpha$ and $\beta$, for each phase, wurtzite and zinc blende. A set of $E_g$ data was calculated for zinc blende AlN at discrete temperature points, from 0K to 420K. Figure \ref{fig:alumnitride} plots the wavelengths calculated using Equation (\ref{eqn:varshni}) in conjunction with Planck's energy equation.  
\begin{table}[!htb]
    \centering
    \caption{Varshni parameters for the two phases of AlN.}
    \begin{tabular}{lcc}
    \hline
     Parameter &  Wurtzite  & Zinc blende \\
    \hline
     $E_{g,0}$ (eV) & 6.23 &  6.23  \\
     $\alpha$ (meV/K)&  1.799 &  0.593  \\
    $\beta$ (K)  & 1462 & 600 \\
     \hline
    \end{tabular}
    \label{tab:alumnitride}
\end{table}
Figure \ref{fig:alumnitride}(b) shows the closeness of fit from 200K. The Pearson's $R^2$-coefficient of 0.994 shows a high degree of linearity above 200K. 
The wavelength variation in Figure \ref{fig:alumnitride}(b) was then used to test the hypotheses presented above, as though they were experimental $\lambda$ versus $T$ data. Figure \ref{fig:comparisons} compares the band gaps and wavelengths calculated using published Varshni coefficients against those calculated using the hypotheses above. At first glance, the results appear very different. However, on closer inspection they can be seen to differ by no more than 2\% over the 200K--420K temperature range. 
\begin{table}[!htb]
    \centering
    \caption{Hypothetically determined Varshni parameters for zinc blende AlN. The absolute error is $\Delta$.}
    \begin{tabular}{llc}
    \hline
     Parameter & Value & $|\Delta|$ (\%)\\
    \hline
     $m$ (nm/K) & 1.888$\times 10^{-2}$ & 0.2 \\
     $\delta$ (/K) & 9.58$\times 10^{-5}$ & 1.4 \\
     $\lambda_0$ (nm) & 196.8 & 1.2\\
     $E_{g,0}$ (eV) & 6.30 & 1.2 \\     
     $\alpha$ (meV/K) & 0.604 & 2.4\\
     $\beta$ (K)  & 746.5 & 24.3\\
     \hline
    \end{tabular}
    \label{tab:alumnitrideresults}
\end{table}
The standard deviation, $\sigma$, in the gradient $m$ determined using $N$ data points was calculated using the graphically determined standard error, $\mu$, in the linearization according to $\sigma=\mu\sqrt{N}$. The error in $\lambda_0$ has been taken as the absolute error relative to the known value of $\lambda_0$ for AlN. Furthermore, since $m$=$k\alpha\lambda_0^2$ where $k$=$e/\varepsilon$, it is readily shown that the maximum uncertainty in $\alpha$ and $\delta$ are:
\begin{equation}
    \Delta\alpha=\alpha\left(\frac{\Delta m}{m}+2\frac{\Delta \lambda_0}{\lambda_0}\right), \quad \Delta\delta=\delta\left(\frac{\Delta m}{m}+\frac{\Delta \lambda_0}{\lambda_0}\right).
\end{equation}
The largest absolute error is observed in $\beta$, at 24.3\%. 
This, however, is expected since there is some variance in the reported literature value. These results show excellent agreement between the model and the expected literature values. The next section presents an experimental setup to test actual LEDs using the same concepts. It differs from the calculations based on the literature in only one respect, namely, that we do actual wavelength measurements on the LEDs to determine the Varshni parameters. 

\subsection{Experimental}

\begin{figure}[!htb]
  \centering     
      \includegraphics[width=0.9\textwidth]{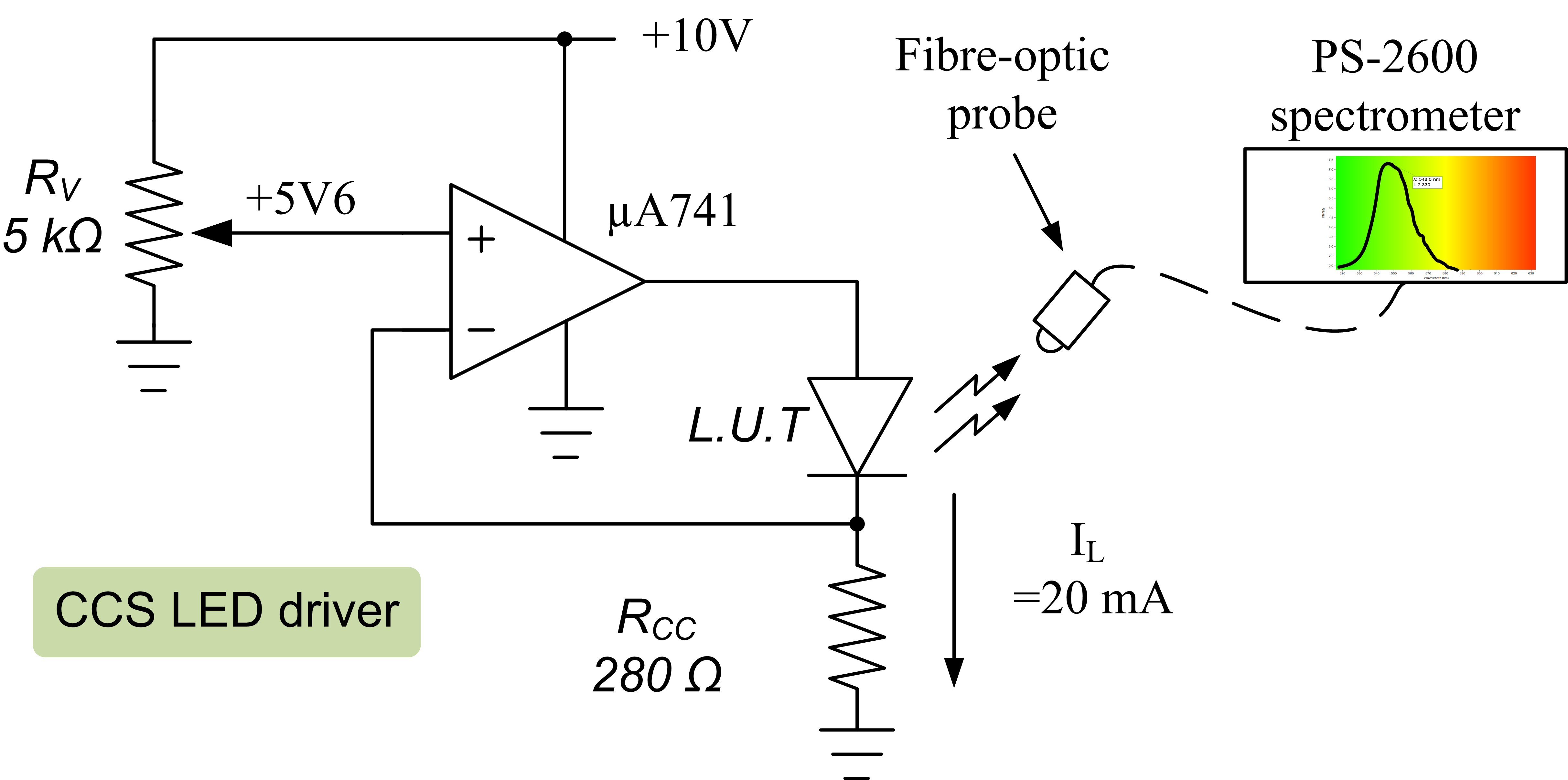}
    \caption{Schematic diagram of the constant-current LED driver and emission detection using a USB spectrometer sensed using an optical fibre probe. LUT is the LED under test.}
    \label{fig:ccsource}
 \end{figure}

\begin{figure}[!htb]
  \centering     
      \includegraphics[width=0.9\textwidth]{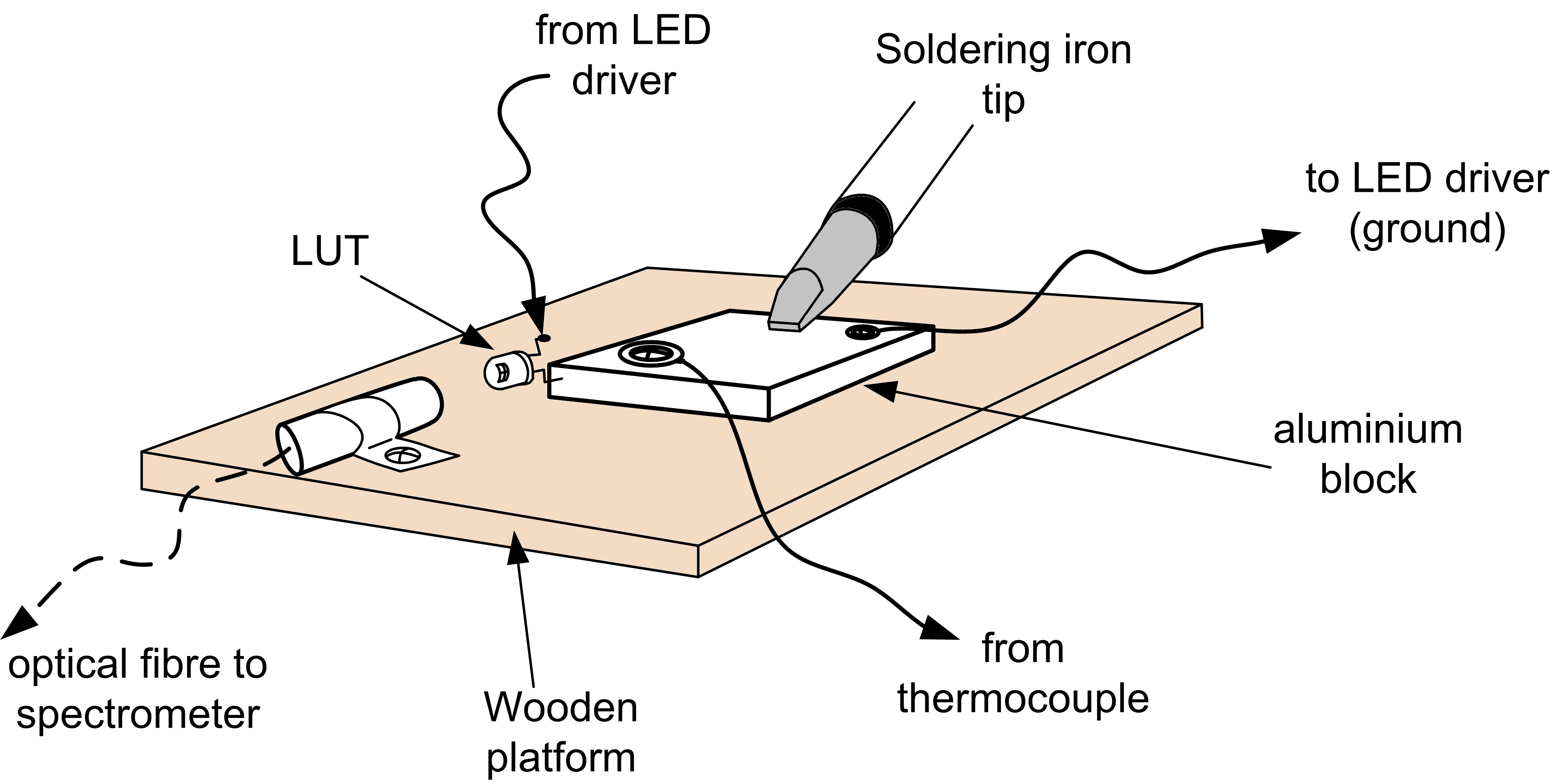}
    \caption{A diagram showing the relative mounts of LEDs, fibre optic probe, thermocouple sensor and heat source.}
    \label{fig:mount}
 \end{figure}

Actual measurements were are done on five commercially sourced LEDs: one ultraviolet, three visible (blue, green, red) and one infrared. Figure \ref{fig:ccsource} shows the basic operational amplifier constant-current source (CCS) used to bias all the LEDs. There are advantages to using this circuit. First, the CCS responsively sets the correct, unique LED turn-on voltage on its $I-V$ characteristic at the LED current (e.g. 20mA). The current is preset through $R_{CC}$ and $R_V$. Second, the low constant-current keeps the LED self-heating, $V_LI_L$, to a mimimum. This is inherently advantageous since the forward voltage $V_L$ of a constant-current biased p-n junction diode decreases linearly with increasing diode temperature, a fact that is exploited in linear, wide-range diode thermometers \cite{ocaya2013linear}. Therefore, the self-heating diminishes as the experiment progresses for each LED. The spectrometer is a PS-2600 USB-connected device from PASCO Scientific with a measurement range of 380 nm to 950 nm \cite{pasco}. Figure \ref{fig:mount} shows how the LED is heated and its temperature monitored. All the components are securely affixed to a solid wooden block. The wooden base also has a high heat capacity that serves to stabilize the temperature rise, thereby preventing rapid temperature variations during heating. This improves the margin of error in the measured LED temperature. Heat is applied using a solder iron tip (preset on its station to 350$^o$C) onto an aluminium plate (1 cm$\times$2 cm$\times$1 mm thick, screwed down onto the wooden base). The heat is conducted into the LED from the aluminium plate. The LED cathode forms the subtrate lead out of the device and is responsible for conducting heat into the LED chip. It was kept as short as was practically possible to maximize its heat conduction. A Chromel-Alumel thermocouple (k-type) was also attached to the aluminium plate to allow direct temperature readout in $^o$C on a digital multimeter. The thermocouple and the LED are mounted in close proximity to each other. Precautions were taken to avoid direct heating of LED/thermocouple by the soldering iron.
These measurements can be automated if a data-acquisition is available. 
 
\section{Results and Analysis}
\begin{figure}[!htb]
 \begin{subfigure}{0.5\textwidth}
\includegraphics[width=\linewidth, height=5cm]{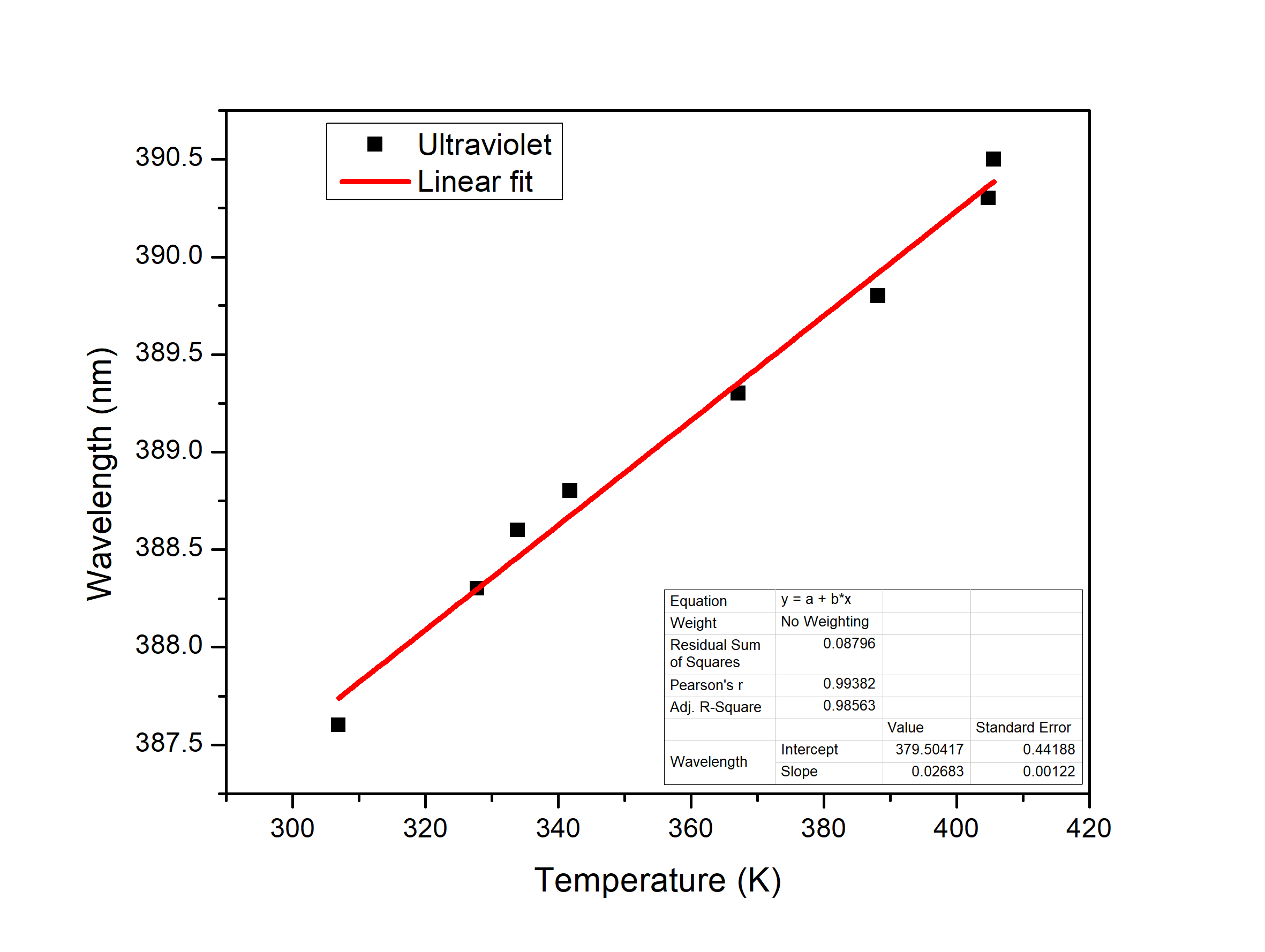} 
\caption{Ultraviolet}
\label{fig:uv}
\end{subfigure}
\begin{subfigure}{0.5\textwidth}
\includegraphics[width=\linewidth, height=5cm]{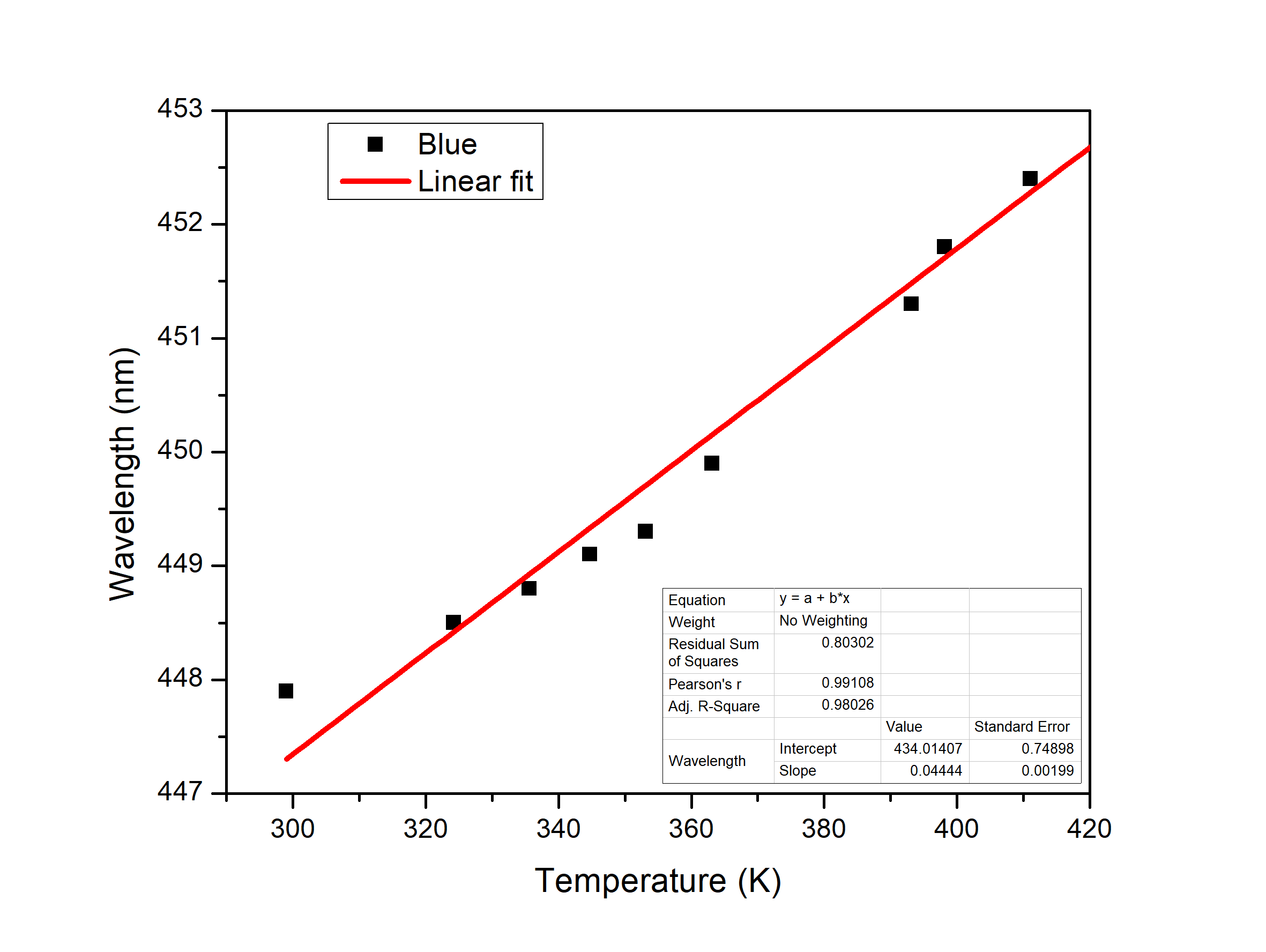}
\caption{Blue}
\label{fig:blue}
\end{subfigure}
 \begin{subfigure}{0.5\textwidth}
\includegraphics[width=\linewidth, height=5cm]{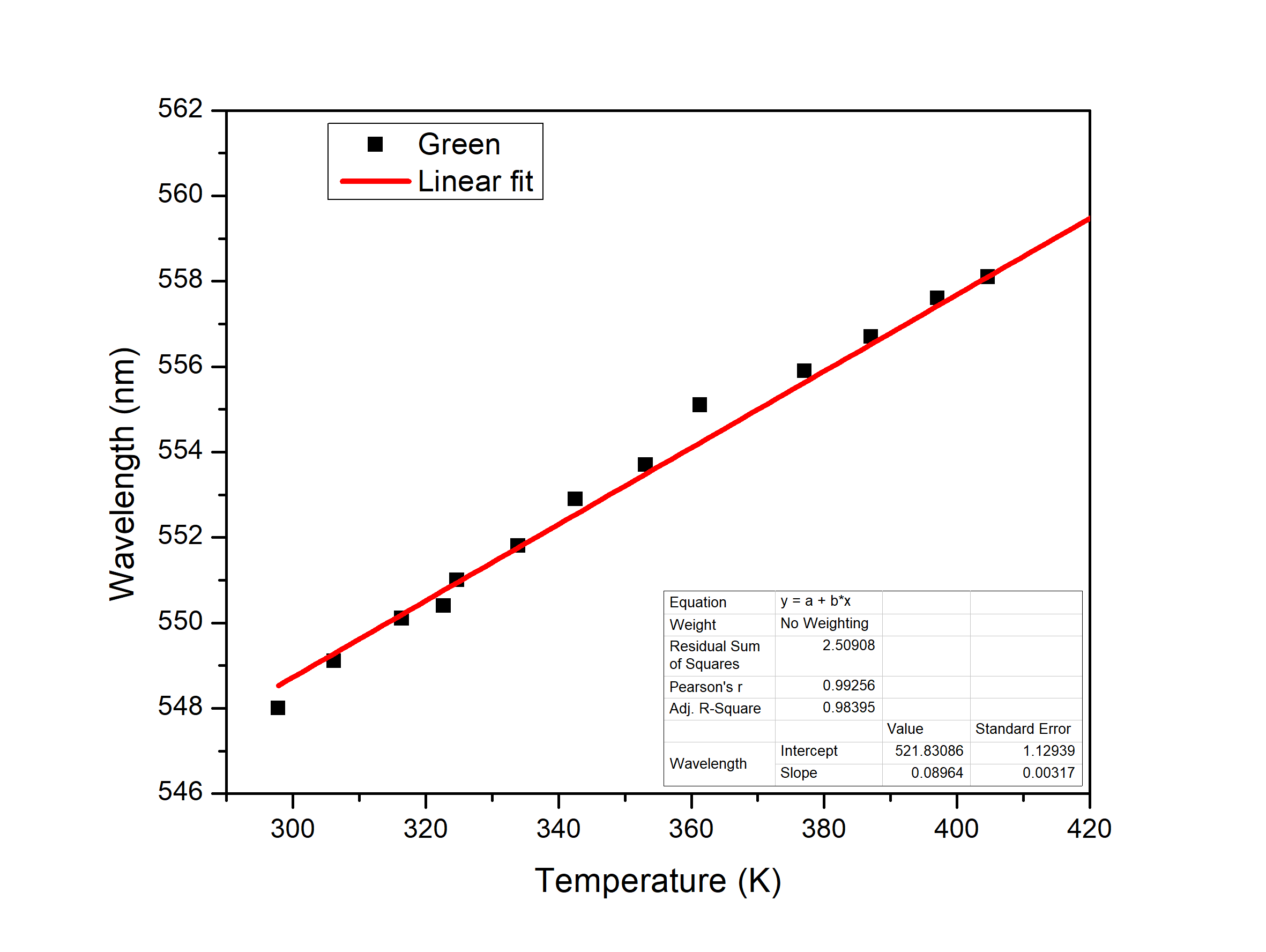} 
\caption{Green}
\label{fig:green}
\end{subfigure}
\begin{subfigure}{0.5\textwidth}
\includegraphics[width=\linewidth, height=5cm]{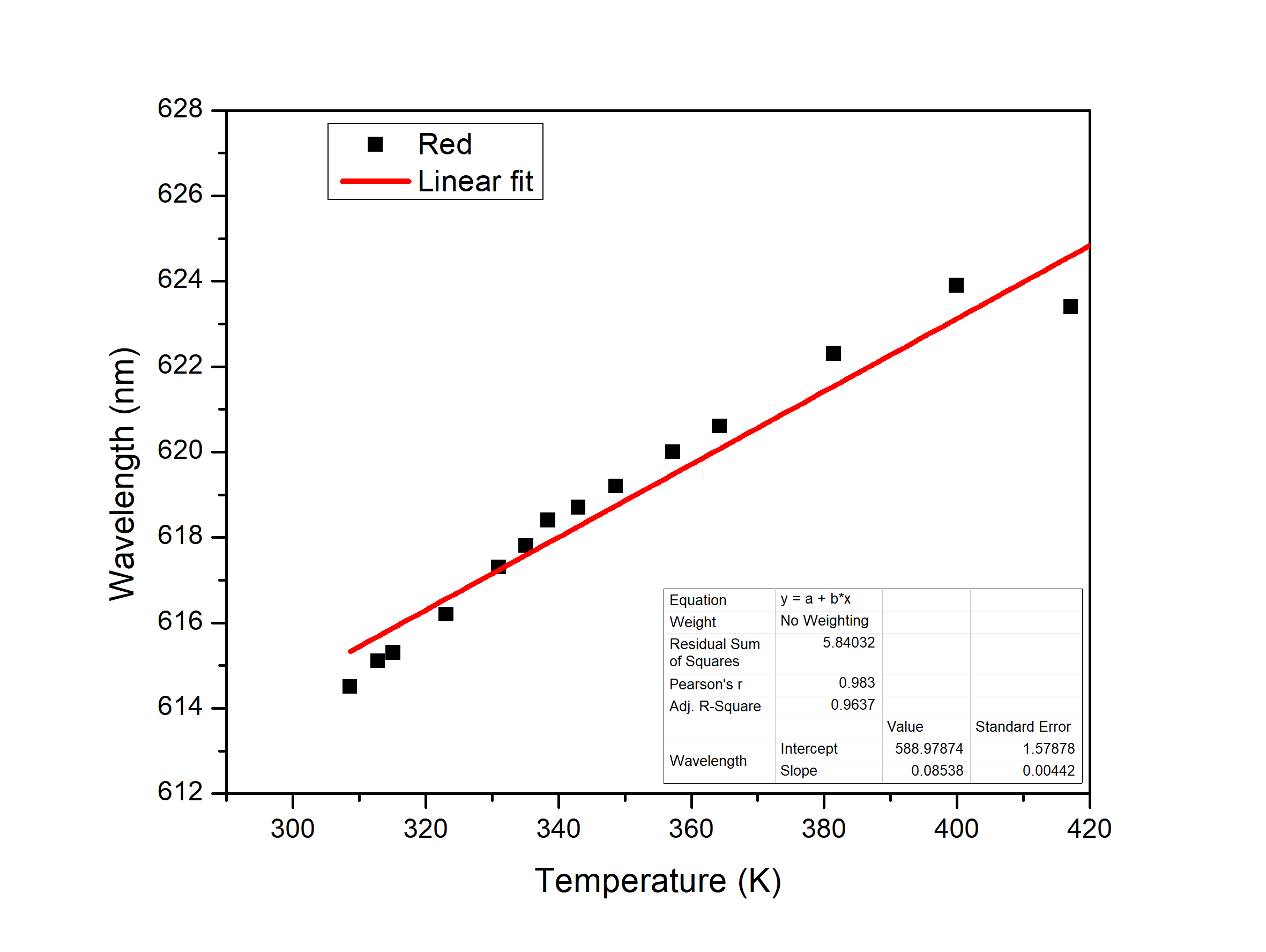}
\caption{Red}
\label{fig:red}
\end{subfigure}
\begin{subfigure}{0.5\textwidth}\centering
\includegraphics[width=\linewidth, height=5cm]{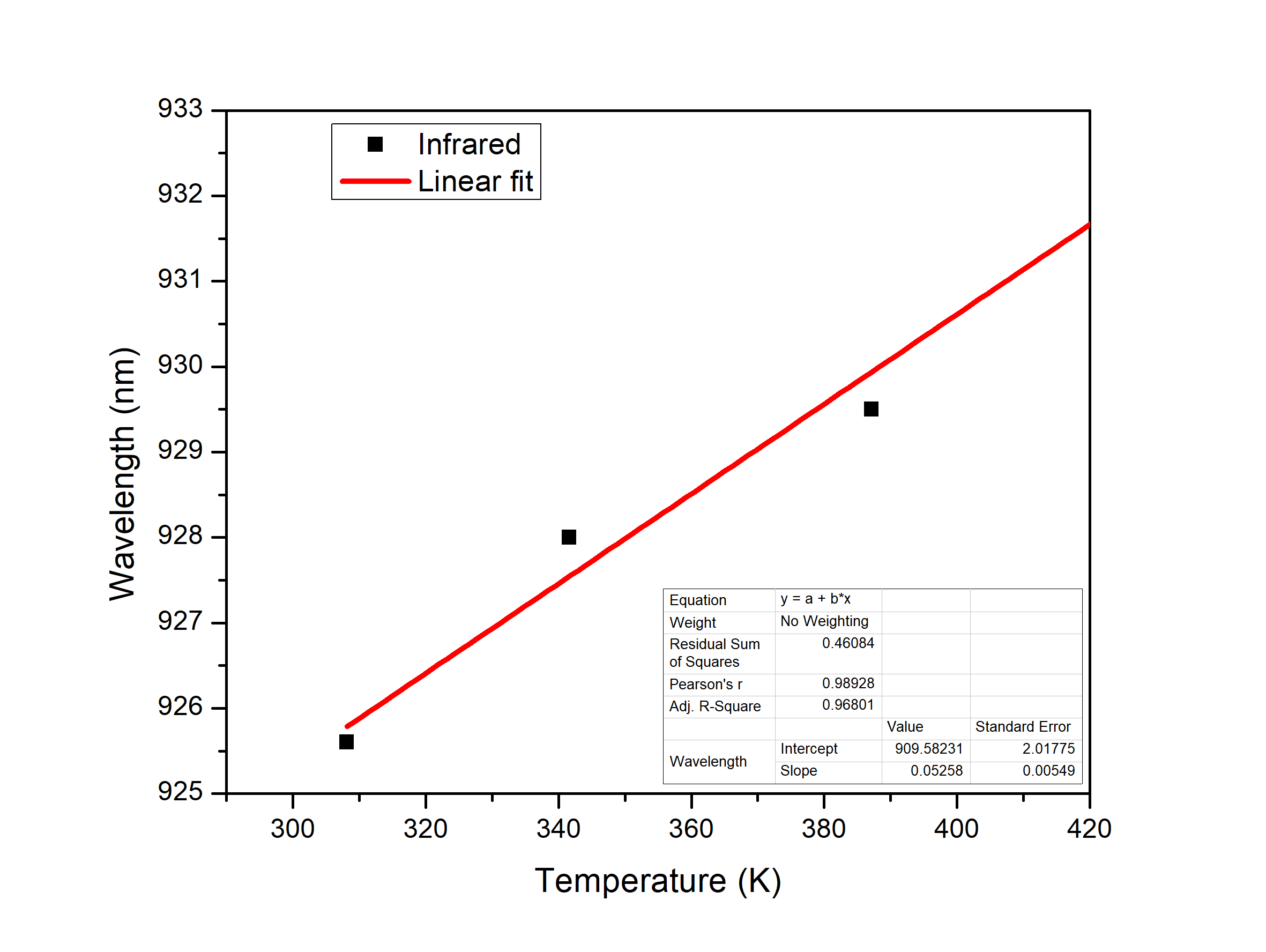} 
\caption{Infrared}
\label{fig:ir}
\end{subfigure}
\caption{Experimentally determined wavelength versus temperature variations for five LEDs. The measurement range of the spectrometer is 380--950 nm.}
\label{fig:LEDresults}
\end{figure}

The following results are based on actual measurements. The upper limit of the spectrometer is 950 nm. This limited the infrared measurements to just four points. However, the infrared measurements follow the observed linear trends of the other LEDs. 
Figures \ref{fig:LEDresults}(a)-(e) show the experimentally determined $\lambda$-$T$ variations.
Table \ref{tab:LEDresults} shows the calculated LED parameters, based on the foregoing hypotheses. 
\begin{table}[!htb]
    \centering
    \caption{Varshni parameters determined for different LEDs using the hypothesis. The LED materials were not known before hand.}
    \begin{tabular}{llllll}
    \hline
     Parameter & UV & Blue & Green & Red & IR \\
    \hline
     $m$ ($\times 10^{-2}$ nm/K) & 2.6830 & 4.4440 & 8.9640 & 8.5380 & 5.2580 \\
     $\delta$ ($\times 10^{-4}$/K) & 0.7761 & 0.9723 & 1.6490 & 1.4700 & 0.6126 \\
     $\lambda_0$ (nm) & 378.6 & 435.2 & 522.3 & 587.8 & 908.5 \\
     $E_{g,0}$ (eV) & 3.28 & 2.85 & 2.38 & 2.11 & 1.37 \\     
     $\alpha$ (meV/K) & 0.2544 & 0.2772 & 0.3918 & 0.3103 & 0.8367 \\
     $\beta$ (K) & 875.2 & 915.9 & 908.5 & 942.1 & 893.8 \\
     \hline
     {\bf\tiny ($\lambda,T$) data sets used for $\delta$, $\lambda_0$ and $\beta$:}\\
     $T_1$ (K) & 306.95 & 299.05 & 297.85 & 308.65 & 308.15 \\
     $T_2$ (K) & 405.65 & 445.55 & 427.85 & 446.35 & 423.15 \\
     $\lambda_1$ (nm) & 387.60 & 447.90 & 548.00 & 614.50 & 925.60 \\
     $\lambda_2$ (nm) & 390.50 & 454.10 & 559.20 & 626.40 & 932.00 \\
     \hline
    \end{tabular}
    \label{tab:LEDresults}
\end{table}

\section{Conclusions}

We have suggested a novel method that, presuming the validity of Varshni's relation, allows the linear tuning of the wavelength of a LED through its temperature. We formulate and prove two hypotheses and devise a basic experimental setup to test them. The experimental results on the LEDs tested show that the wavelengths varied by up to 12 nm for the green/red LEDs over the experimental temperature range. The change in wavelength is reversible on cooling and was visually discernible for the green and red LEDs. Thus, we add a measurement tool that has the advantage of determining the Varshni coefficients alongside the 0K band gap through simple measurements at room temperatures and above. The error analyses presented suggest that the calculations are accurate to a remarkable degree, with many of the uncertainties being less than 2\%. The method is non-invasive and does not operate the LED beyond its maximum electrical operating points. The wavelength changes are therefore reversible. The approach could prove useful for the testing of existing LEDs that are subjected to external influences, such as high radiation fluence, to study the impact on parameters related to the bandgap in a practical setting. Finally, since we have established that LEDs have low wavelength stability over temperature, this work could form the basis of a control strategy to achieve wavelength stability for more specific applications. 

\appendix


\bibliographystyle{elsarticle-num}

\end{document}